\documentclass[manuscript]{acmart}

\PassOptionsToPackage{hyphens}{url}
\usepackage{url}

\copyrightyear{2022}
\acmYear{2022}
\setcopyright{rightsretained}
\acmConference[HT '22]{Proceedings of the 33rd ACM Conference on Hypertext and Social Media}{June 28-July 1, 2022}{Barcelona, Spain}
\acmBooktitle{Proceedings of the 33rd ACM Conference on Hypertext and Social Media (HT '22), June 28-July 1, 2022, Barcelona,
Spain}\acmDOI{10.1145/3511095.3536373}
\acmISBN{978-1-4503-9233-4/22/06}

\begin{document}

\title[Hyperownership]{Hyperownership: Beyond the Current State of Interaction with Digital Property}

\author{Amaury Trujillo}
\email{amaury.trujillo@iit.cnr.it}
\orcid{0000-0001-6227-0944}
\affiliation{%
    \institution{IIT-CNR}
    \streetaddress{Via G. Moruzzi 1}
    \city{Pisa}
    \country{Italy}
    \postcode{56124}
}

\begin{abstract}
The introduction of novel technology has oftentimes changed the concept of ownership. Non-fungible tokens are a recent example, as they allow a decentralized way to generate and verify proof of ownership via distributed ledger technology. Despite crucial uncertainties, these tokens have generated great enthusiasm for the future of digital property and its surrounding economy. In this regard, I think there is an untapped opportunity in applying a hypertext approach to augment such highly structured ownership-based associations. To this end, in this work I propose \emph{hyperownership}, based on the premises that property is the law of lists and ledgers, and that hypertext is an apt method to inquiry such a ledger system. In spite of the significant risks and challenges to realize such a vision, I believe that it has great potential to transform the way with which we interact with digital property.
\end{abstract}


\maketitle

\section{Introduction}

Ownership can be described, in very broad terms, as the set of exclusive rights over property, usually classified as tangible (e.g., real state, chattel) or intangible (e.g., intellectual property, digital objects).
Such rights are universally recognized as fundamental~\cite{akkermans2018comparative}, yet the very concept of ownership evolves over time, according to the mores of society and technological innovations of the place and moment. Ownership has in fact been intrinsically linked to information technology (in its broadest sense) from the very beginning of history, as attested by the small clay objects inscribed with proto-cuneiform dating back to 8000 BCE found in the Near East ---called \emph{tokens}--- used to keep track of the number of animals owned~\cite[p. 22]{daniels1996world}. Millennia later, the printing press, another technological writing milestone, greatly eased the massive copying of books and also gave way to a new form of piracy of intellectual property~\cite[\S1]{johns2010piracy}. More recently, digital technology, which allows perfect copies of digital objects at near-zero cost, brought into question aspects of goods once considered to be central~\cite{liu2000owning}, such as rivalry (impossible simultaneous use) and scarcity (limited availability). Incidentally, such «fundamental shift from tactile to digital, physical to code, and hard to soft media» in ownership is also reflected by hypertext within the history of information technology, as described by \citeauthor{landow2006hypertext}~\cite[pp. 29--41]{landow2006hypertext}.

Moreover, present-day digital ecosystems ---with most based on hypermedia--- have heavily influenced our ownership experience. For instance, whereas traditional market models are based on ownership, the \emph{sharing economy} is based on using and sharing products among each other, and is driven by  consumer behavior (preference for convenience, low prices, sustainability), social networks and electronic markets, and mobile devices and electronic services~\cite{puschmann2016sharing}. Another example is the push towards a \emph{subscription model} of digital goods and services, such as news, music, video, videogames, and apps. Nonetheless, not all  changes in digital ownership are welcome, and much debate is being made regarding the rights of buyers, creators, and distributors of digital objects, and the loss of expected privileges associated with ownership~\cite{liu2000owning}. Further, the increasing practice of license-only digital objects and the abuse of digital rights management (DRM) have given rise to harsh criticism against this “end of ownership” in the digital economy~\cite{perzanowski2016end}.

\section{A new kind of digital ownership}

Discontent with diminishing digital property rights led many tech enthusiasts to seek alternative approaches. Arguably, one of the most interesting is the use of a non-fungible token (NFT) to certify the authenticity and ownership of tangible or intangible assets. An NFT is a unique identifier recorded using distributed ledger  technology (DLT), which is based on decentralized immutable lists of records spread over a peer-to-peer network without the need of a trusted authority, with the most famous example being blockchain~\cite{swartz2017blockchain}. NFTs are a specific form of \emph{smart contracts}, versatile transaction protocols that automatically execute, control, and document an agreement without the need of a trusted intermediator~\cite{antonopoulos2018mastering}. Amusingly enough, NFTs seem to carry on the ancient practice of using \emph{tokens} to keep track of property, only this time via digital cryptographic hashes instead of inscriptions on clay.

In the last couple of years, there has been a surge in public interest on NFTs due to the high-profile sale of certain digital assets~\cite{trujillo2022surge}, such as the sale for US\$5.4M of the initial implementation source code of the World Wide Web by Tim-Berners Lee.\footnote{\url{https://www.bbc.com/news/technology-57666335}} Users and buyers of NFTs are allured by the promise of ownership, traceability, and financial gain~\cite{mackenzie2021nfts}. And the high figures involved in much publicized sales have sparked a frenzy in the development of NFTs applications and ecosystems in different sectors, such as collectibles, sports, visual arts, and video games~\cite{nadini2021mapping}. Even “traditional” social media have begun integrating NFTs. Twitter now can display special NFT profile pictures with a hexagonal shape present in the Ethereum blockchain (the most common DLT for NFTs), and Reddit has established a collaboration with OpenSea (the largest NFT marketplace) to sell unique animated avatars, called CryptoSnoos. Many more ideas are emerging around NFTs and DLT beyond finance, such as decentralized autonomous organizations (DAOs), a novel manner to codify and automatically manage governance via smart contracts without a central leadership~\cite{hassan2021decentralized}. Hence, due to the aforementioned commonalities with respect to information technology, I believe that using an hypertext approach to augment our capabilities on digital ownership based on DLT is a logical yet untapped opportunity.

\section{Envisioning hyperownership}

Bitcoin, the first successful decentralized virtual currency, gave way to the idea of using DLT as a means to register property ownership.
In particular, \citeauthor{fairfield2014bitproperty}~\cite{fairfield2014bitproperty} posited that property could be seen as a set of information describing who may do what, when, and with which resource: «property is the law of lists and ledgers». In this perspective, property does not consist in the \emph{thing} itself, but in the packaging, tracking and transmission of data regarding ownership via information systems; thus, a property system has two key actions: \emph{store} and \emph{communicate} ownership information.

However, DLT has mainly focused on improving the latter action from the point of view of machines, neglecting the human element. For example, in a survey of DLT applications from the perspective of human-computer interaction (HCI), \citeauthor{elsden2018making} state that these «are currently overwhelmingly driven by a mix of engineering, investment and crypto-anarchist visions», and they express the need for more research on the human challenges of DLT, such as design for trust, algorithmic governance, influence on societal and individual values, ease of use for end users, and implications of publicly sharing one's property data~\cite{elsden2018making}. \citeauthor{glomann2019improving} further confirm the need of a human-centered approach given crucial DLT issues: high onboarding learnability; lack of design considerations for usability, efficiency, and accessibility; and design challenges in decentralized applications~\cite{glomann2019improving}.
In this respect, I believe there is an unexploited opportunity in using a hypertext approach to tackle these issues, for as hypertext has been historically touted as a means to augment human intellect~\cite{van2019reflections}, by allowing a person to approach complex problems, to fulfill their comprehension needs, and to derive solutions to problems~\cite{engelbart1962augmenting}. To seize such an opportunity, I think the view posited by \citeauthor{atzenbeck2019hypertext}~\cite{atzenbeck2019hypertext} of a «hypertext method of inquiry, as way of viewing arbitrary systems», is particularly apt. In this view, hypertext shifts its focus from a particular set of technologies and techniques, to the more general explicit association of information. \emph{Hyperownership} is thus a concept to augment human capabilities on digital ownership, by synthesizing these two ideas: property is the law of lists and ledgers, and hypertext is a method of inquiry of arbitrary systems.

Based on their form of structure from a hypertext perspective~\cite{atzenbeck2019hypertext}, NFTs are a first class structure, as by definition DLT records explicit relationships between resources and allows the association of metadata with a link for a given transaction. On top of these records, we could use a new representation to exploit the explicit ownership associations in new ways. For instance, we could use \emph{temporal bipartite graphs} to represent the evolving relationships between the two main kinds of DLT entities: tokens (property) and their respective account (owner). Compared to their static and unipartite counterparts, temporal bipartite graphs model better the complexity and dynamism of real-world problems~\cite{chen2021efficiently}. In this \emph{owner--property} temporal bipartite graph, the state of ownership in a given moment could be computed from a derived static graph, i.e., a \emph{snapshot}, in which each token is associated with at most one owner and each owner is associated with zero or more tokens. A second ownership-based useful representation could be a \emph{property--usage} temporal bipartite graph, in which we model distinct usages of property tokens in a given context (e.g., profile picture in social media, an audio or video reproduction). In the snapshots of this graph, each property could be associated to zero or more usages. Finally, a third temporal bipartite graph, \emph{owner--usage} could be derived from the first two. Further styling could be done through automatic context-aware hypermedia sculpting to remove irrelevant links based on context or manual calligraphic linking by users~\cite{hargood2016patterns}.

\begin{figure*}
\includegraphics{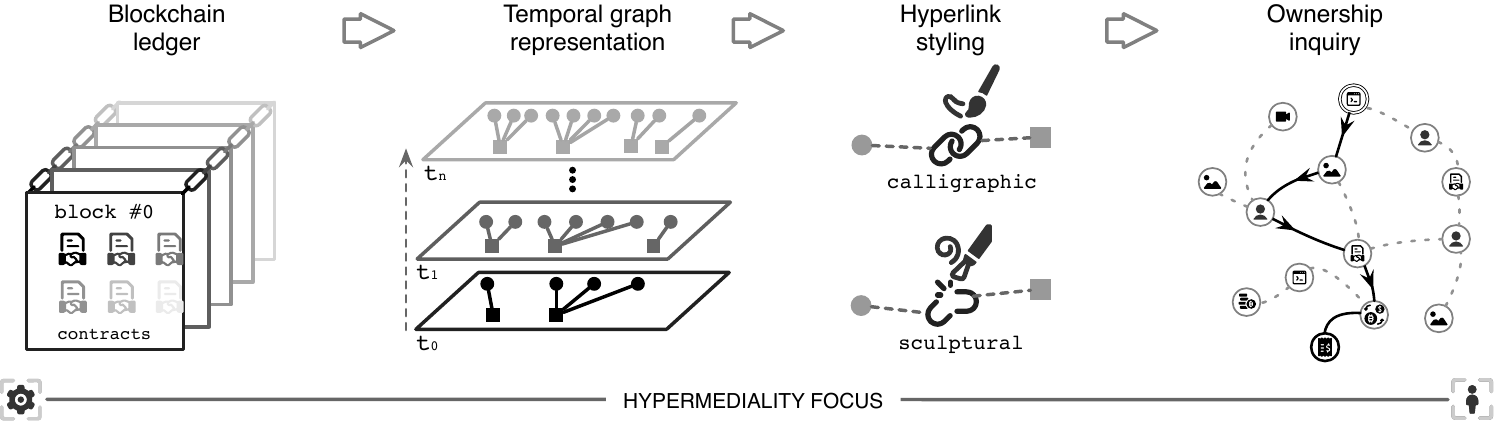}
\caption{Conceptual diagram of applying a hypermedia approach to blockchain so as to inquiry on the ownership of digital objects, in which there is an increasing shift in focus from the machine to the human aspects of the system.}
\label{fig:diagram}
\Description{The figure is divided into four sections. In the first there is a blockchain containing a series of smart contracts. In the second, these are converted to temporal bipartite graphs. In the third, further customization is illustrated via calligraphic and sculpture hypermedia styles to respectively increase or reduce existing links. In the final section, an example of ownership inquiry is shown, in which the associative trail between a usage of a given asset and its monetary cost, passing through the related seller, owner, and cryptocurrency exchange.}
\end{figure*}

With such an approach, illustrated in Fig.~\ref{fig:diagram}, we could augment human capabilities to inquiry on ownership by: using novel HCI techniques to visualize and interact with digital property; leveraging recent advances in big graph mining~\cite{aridhi2016big} to discover interesting phenomena and gain insight into the dynamics of ownership; integrating existing knowledge graphs~\cite{ji2021survey} to enrich the semantics of the ownership network and viceversa; creating novel non-linear narratives regarding the history of digital objects and the preferences of owner accounts; easing the exploration and recommendation of potentially interesting digital objects to buy, use, or admire; and much more.

\section{Risks and challenges}

Due to the uncertainties surrounding NFTs, there are significant risks in undertaking the realization of hyperownership. For instance, despite the promise and eagerness of improved usage and financial control and financial by digital content creators, many are divided on the merits of NFTs~\cite{kugler2021non}.
Above all, crucial issues remain unanswered regarding the ownership of NFTs and the objects that they represent~\cite{lydiate2021crypto}, such as a likely economic bubble, legal recognition, and the currently rising fraud and stealing in the community; as simply put by \citeauthor{joselit2021nfts}~\cite{joselit2021nfts}: «The NFT is a social contract that values property over material experience. That contract can be broken.» Nevertheless, I think we should  look beyond these hectic times and focus on the innovative technology and techniques of NFTs and DLT in general. After all, the hypertext community has already gone trough a similar tumultuous technological period caused by hype and unwise investments, during the rise of the World Wide Web and subsequent \emph{dot-com bubble burst}~\cite{wheale2003bursting}.

Many technical and ethical challenges must also be resolved to realize hyperownership. For instance, at the moment we have competing DLT and NFT community-driven standards, resulting in low cross-platform compatibility, albeit many initiatives are in course to remedy this~\cite{williams2020cross}. In addition, DLT and related technologies present significant challenges in software engineering~\cite{wessling2018engineering}. For instance, many DLT applications store assets on the InterPlanetary File System (IPFS), a distributed peer-to-peer hypermedia protocol aimed to be resilient and persistent~\cite{huang2020blockchain}, which entails a different development approach. In fact, given this shift in paradigm, many people call Web3 a potential decentralized and token-based iteration of the Web~\cite{voshmgir2020token}, albeit this has been decried as a mere buzzword~\cite{newitz2022web3}.
Still, Web3 tenets might represent the next step in hypermedia infrastructure, which began as monolithic systems, then client-server systems, later as open hypermedia systems (OHS), and currently as component-based OHS~\cite{atzenbeck2017revisiting}. Hence, distributed and decentralized component-based OHS (using for instance IPFS for media storage) should be further explored. 
Another significant technical challenge concerns scalability, especially of a graph representation of DLT records which might manifest rapid exponential growth~\cite{alabi2017digital}. In this regard, the community on big graph processing systems is working towards adapting new distributed workloads, standard models, and suitable performance metrics to model complex real-world phenomena, with a keen eye towards the possibility of creating a Big Graph Memex~\cite{sakr2021future}, inspired by Bush's original Memex concept. Hyperownership as described herein thus fits and builds upon such an ambitious vision.

Last but not least, ethical issues are also significant for hyperownership. Privacy, a main driver of DLT~\cite{hassan2019privacy}, is far from perfect, and as some of us can imagine, not everyone is keen in facilitating access to the information regarding what we own. Such issues are also present in hypertext, given that despite its potential to empower users and democratize access to information, it is susceptible to surveillance and manipulation~\cite[Ch. 8]{landow2006hypertext}. Indeed, both sculptural and calligraphic hypertext styles could be used to restrict access or to manipulate the narrative regarding the ownership of certain assets by certain individuals or groups. In fact, it could argued that both hypertext and DLT influence and are influenced by political motives, as respectively demonstrated by the OHS and DAOs. Hence, particularly attention should be paid to these sensitive issues in order to avoid abuse, intentional or not.

\section{Conclusion}

Once again, novel technology is changing the perception and experience of ownership, more so in an environment as \emph{interwingled} as the digital economy. This change represents an opportunity for the development of novel hypertext ideas, such as hyperownership, to unravel and augment our  relationship with digital property. Hence, I invite the hypertext community to look and take  inspiration by the novel (yet ancient) \emph{token economy}.

\bibliographystyle{ACM-Reference-Format}
\bibliography{references}
\end{document}